# Room-temperature electric field effect and carrier-type inversion in graphene films

K.S. Novoselov[1], A.K. Geim[1], S.V. Morozov[2], S.V. Dubonos[2], Y. Zhang[1], D. Jiang[1]

[1]Department of Physics, University of Manchester, M13 9PL, Manchester, UK

[2]Institute for Microelectronics Technology, 142432 Chernogolovka, Russia

**The ability to control electronic properties of a material by externally applied voltage is at the heart of modern electronics. In many cases, it is the so-called electric field effect that allows one to vary the carrier concentration in a semiconductor device and, consequently, change an electric current through it. As the semiconductor industry is nearing the limits of performance improvements for the current technologies dominated by silicon, there is a constant search for new, non-traditional materials whose properties can be controlled by electric field. Most notable examples of such materials developed recently are organic conductors [1], oxides near a superconducting or magnetic phase transition [2] and carbon nanotubes [3-5]. Here, we describe another system of this kind – thin monocrystalline films of graphite – which exhibits a pronounced electric field effect, such that charge carriers can be turned into either electrons or holes. The films remain metallic, continuous and of high quality down to a few atomic layers in thickness. The demonstrated ease of preparing such films of nearly macroscopic sizes and of their processing by standard microfabrication techniques, combined with submicron-scale ballistic transport even at room temperature, offer a new two-dimensional system controllable by electric-field doping and provide a realistic promise of device applications.**

The surface charge that can be induced in a material by the field effect is limited by electric breakdown of dielectric materials ($\approx 10^9$ V/m) and cannot normally exceed concentrations of $\approx 10^{13}$ cm$^{-2}$. Such concentrations are sufficient to change a thin semiconducting layer from an insulator to a metal, the effect widely used in various device applications such as, e.g., field effect transistors. However tempting, the use of the field effect cannot be extended to normal metals (e.g., to develop all-metal transistors) because of high carrier densities even for the case of mono-atomic layers ($\approx 10^{15}$ cm$^{-2}$). A possible exception is semimetals, which are distinct from normal metals because of their low carrier concentrations, close to those characteristic for doped semiconductors. In semimetals, an electric field effect can generally be expected for layers up to several nm thick (in thicker samples, any changes in electronic properties would inevitably be obscured by the bulk conductance unaffected by electric field). Unfortunately, so thin metallic films are mostly unstable and become discontinuous, which so far has not allowed development of semimetal devices controllable by the field effect.

In this report, we focus on graphite, which stands out from other semimetals due to its layered crystal structure, stable even at temperatures $\approx 3500°$C. Within each layer, carbon atoms are densely packed in a benzene-ring structure with a nearest-neighbour distance of $\approx 1.4$Å. The layers (often referred to as graphene) are relatively loosely stacked on top of each other with an interlayer distance of $\approx 3.35$Å. This leads to mechanical and electrical properties of graphite being highly anisotropic. Graphite has a small density of charge carriers ($\approx 1$ carrier per $10^4$ carbon atoms) and equal concentrations of electrons and holes ($N_e \approx N_h \approx 5 \cdot 10^{18}$ cm$^{-3}$ at 300K) [6-8]. At room temperature, the carriers typically exhibit in-plane mobility μ of $\approx 15,000$ cm$^2$/V·s [6,7] and, reportedly [9,10], even as high as $\approx 100,000$ cm$^2$/V·s. At liquid-helium temperatures, μ can routinely reach $10^6$ cm$^2$/V·s [6-10]. For comparison, room-temperature mobilities in Si and GaAs are $\approx 1,000$ and $\approx 9,000$ cm$^2$/V·s, respectively (μ is known to be one of the most important parameters for semiconducting device applications, determining their frequency response and power consumption).

We have found a way to make monocrystalline films of graphite with thickness $d$ down to a few nm, process them into transistor-like devices and control their properties by electric field. The films are made by repeated peeling of small mesas of highly-oriented pyrolytic graphite (HOPG). This procedure, described in Methods, has already allowed us to controllably prepare films as thin as ≈15Å (i.e. only 5 graphene layers thick) and as large as a hundred μm across (Fig. 1). The graphene films remain metallic and, as an indicator of their high quality, show no significant reduction in μ as compared to the mobility of our original HOPG (≈15,000 cm$^2$/V·s). To the best of our knowledge, no other film of similar thickness remains continuous and metallic under ambient conditions. One might find it a useful analogy to compare the prepared films with carbon nanotubes that are widely considered as a possible base material for future electronics [11]. Carbon nanotubes are often thought of as graphene sheets wrapped up in nm-sized cylinders. Visa versa, our graphene films can be viewed as if they were effectively made from (multi-wall) carbon nanotubes that are unfolded and stitched together to cover a macroscopic area.

To study their electronic properties, the described graphene films were processed into multi-terminal Hall bar devices placed on top of an oxidized Si substrate (Fig. 2a&b). Using both low-frequency lock-in and dc measurement techniques and applying a gate voltage to the substrate, we have studied more than 20 different devices, concentrating on "relatively thick" films ($d \geq 3$ nm), as these were found to exhibit consistent and reproducible behaviour and a room-temperature mobility between 7,000 and 15,000 cm$^2$/V·s. These films are also thin enough to allow major changes in their carrier concentration under relatively modest electric fields of <0.3V/nm, easily achievable for SiO$_2$. Indeed, the surface charge density induced by gate voltage $V_g$ can be estimated as $n = \varepsilon_0 \varepsilon V_g / te$ where $\varepsilon_0$ and $\varepsilon$ are permittivities of free space and SiO$_2$, respectively, $e$ is the electron charge and $t = 300$ nm is the thickness of SiO$_2$. This yields $n/V_g \approx 7.2 \cdot 10^{10}$ cm$^{-2}$/V, which for the case of 3-nm films at 100V exceeds the intrinsic area densities of charge carriers $n_i = N_e \cdot d = N_h \cdot d \approx 1.5 \cdot 10^{12}$ cm$^{-2}$ by a factor of 5.

Figure 2 shows a typical dependence of resistivity $R$ of our graphene films on gate voltage ($R = \rho/d$ is the film resistivity per □, and $\rho$ is the bulk resistivity). On can see that the resistivity changes by a factor of 4 with varying $V_g$. Measurements in magnetic field $B$ revealed that changes in Hall resistivity $R_{xy}$ were even more dramatic, including a sharp reversal of the sign of the Hall effect (Fig. 3). The observed behaviour yields (see below) that the electric-field doping can transform films of this nominally compensated ($N_e = N_h$) semimetal into either completely electron or completely hole conductors. For thicker films, the field-effect-induced changes in $R$ were found to decrease approximately as $1/d$, in agreement with a simple estimate, while changes in $R_{xy}$ became limited to an increasingly narrower region of the field-effect behaviour exhibited by the thinnest samples (Fig. 3b,c). We note that, at zero $V_g$, some of our films were found in the compensated state, as is bulk graphite (Fig. 3b,c), while others – especially the thinner devices – were strongly doped by either donors or acceptors (Figs.2&3). The origin of this unintentional doping remains unclear and can be attributed to either surface defects [12] or absorption of atmospheric gases [13].

The observed changes in electronic properties of graphene films induced by gate voltage can be understood on the basis of the standard two-band model of graphite [6,7]. Fig. 2c shows the results of our calculations for the field effect, using the conventional formula $R^{-1} = d \cdot [e \cdot \mu_e \cdot n_e(V_g) + e \cdot \mu_h \cdot n_h(V_g)]$ where $\mu_e$ and $\mu_h$ are hole and electron mobilities, respectively. The calculated curve is asymmetric with respect to zero voltage, which reflects the difference in electron and hole densities of states due to their different masses $\approx 0.05$ and $\approx 0.04 m_0$, respectively ($m_0$ is the free electron mass) [8]. For low gate voltages, $R$ changes linearly with $V_g$, indicating the substitution of one type of charge carriers by another. At higher $V_g$, such that $n > n_i$, only one type of carriers is left and, accordingly, the resistivity follows the standard dependence $R = (d \cdot ne\mu)^{-1}$ for both negative and positive biases. This leads to the characteristic cusp-like shape of $R(V_g)$-curves. The initial (unintentional) doping results in a shift of the cusp region along the voltage axis (Fig. 2). Similarly, the behaviour of the Hall effect is well described by the same model [6,7] (see Fig. 3a). Three regions of electric-field doping (electron, hole and mixed) are clearly seen on the measured curves. Close to the compensated state, the Hall coefficient $R_H$ depends linearly on $V_g$ and reverses its sign, reflecting the fact that in the mixed-carrier regime $R_H$ is approximately proportional to the difference between electron and hole concentrations. For the voltage regions with only electrons and holes left, $R_H$ decreases with increasing the carrier concentration in a usual way, as $1/n$.

Remarkably, although being only several graphene layers thick, our films exhibited pronounced quantum oscillations in both $R_{xy}$ and longitudinal resistivity $R_{xx}$. This serves as yet another indicator of the quality of the experimental system (our best devices showed $\mu \approx 50,000$ cm$^2$/V·s at 4K). Moreover, the oscillations could be radically altered by gate voltage (see Fig. 4), confirming the fundamental changes in carrier type and concentrations discussed above. The particular film in Fig.4 was initially (at $V_g = 0$) strongly donor-doped and, accordingly, under negative (positive) $V_g$ the Shubnikov-de Haas (ShdH) oscillations became much slower (much faster). Alternatively, the ShdH oscillations could be observed when we fixed field $B$ and varied the gate voltage (Fig. 4a). Similar results were found for samples unintentionally doped with acceptors, although the oscillations were slightly less pronounced. For initially compensated films, we have seen a clear transition between ShdH oscillations due to holes and electrons.

Because of low carrier concentrations, one can generally expect the transport properties of nm-thick graphene films to be two-dimensional (2D) for the whole range of gate voltages. To verify this, we carried out the standard test and measured ShdH oscillations for various angles θ between the magnetic field and the graphene films. The oscillations were found to depend only on the perpendicular component of the magnetic field $B·\cos\theta$, as expected for a 2D system. However, we often found more than one set of ShdH oscillations, which indicates that several 2D subbands could be occupied. For example, one can discern two sets of ShdH oscillations for the 5-nm film in Fig.4. They both change with $V_g$ but changes in the carrier density calculated from the oscillations' frequencies could account only for a part of the charge density induced by gate voltage. In this respect, measurements such as in Fig. 4a provide a better way of determining the number of 2D gases contributing to the conductance. Here, every period ($\delta V_g \approx 12V$) corresponds to changes in the electron density $\delta n = \varepsilon_0 \varepsilon \delta V_g / te \approx 8.6·10^{11} cm^{-2}$ while it requires $2.9·10^{11}$ cm$^{-2}$ electrons ($k = 2eB/h$) to fill a single spin-degenerate Landau level in a 2D gas at 6T. This implies three ($\delta n/k$) types of 2D carriers present in this film.

In conclusion, the described findings open new avenues in terms of both fundamental research and applications. First, one should expect in graphene films the whole wealth of phenomena usually associated with the reduced dimensionality, which have been intensively studied using 2D systems based on semiconducting materials. Furthermore, graphene films can be used to address - from a different perspective - the multitude of phenomena expected in other carbon-based systems such as nanotubes. As for possible applications, further work would undoubtedly lead to the films of larger lateral sizes required for their manufacturability, improvement in quality and a further reduction in thickness (possibly, to the limit of a zero-gap semiconductor expected for a single graphene sheet [6]). The extended analogies with semiconducting and carbon-nanotube devices make us confident that graphene-based electronic components such as field-effect transistors, chemical sensors and ferromagnetic, superconducting and other hybrid devices are already within close reach.

METHODS

**Preparation of graphene films.** 1-mm-thick platelets of a high-grade HOPG were the starting material. We used commercially available HOPG (grades ZYA from *NT-MDT* and HOPG-1 from *SPI* with µ >100,000 cm$^2$/V·s at 4K). Using dry etching in oxygen plasma [14], we first prepared 5 µm-deep mesas on top of the platelets (mesas were squares of various sizes from 20 µm to 2 mm). The structured surface was then pressed against a 1-µm-thick layer of a fresh wet photoresist spun over a glass substrate. After baking, the mesas became attached to the photoresist layer, which allowed us to cleave them off the rest of the HOPG sample. Thin flakes of graphite were repeatedly peeled off the mesas using scotch tape. Then, we selected those flakes that were so thin as to remain practically invisible under an optical microscope. They were released by dissolving scotch's visco-elastic layer in acetone and captured on a Si wafer. Large flakes often appeared crumpled. At this stage, we used ultra-sound cleaning in propanol to remove the crumpled areas as well as occasional thick pieces of graphite. Thin flakes ($d < 10$ nm) were found to attach strongly to SiO$_2$, presumably due to van der Waals and/or capillary forces and, as a result, this left us with thin flat films of relatively large sizes (up to 100 µm) with only a few or no wrinkles (Fig. 1). The thickness of the obtained films was measured by AFM. As a substrate, we used n$^+$-doped Si with a SiO$_2$ layer on top (Fig.2a). To avoid accidental damage (especially, during plasma etching) we chose to use thick (300 nm) SiO$_2$, which necessitated high $V_g$ (up to ±100V) used in the experiments, limited by the electric breakdown of SiO$_2$ (<1V/nm).

**Device microfabrication.** First, by using electron-beam lithography, we fabricated an Al mask on top of a chosen area of a graphene film. This was followed by dry etching in oxygen plasma, which removed graphite from everywhere but underneath the mask. The mask was removed in a weak solution of KOH. Finally, a set of Au contacts was deposited on top of the graphene mesas to allow electrical measurements. These gold-to-graphene contacts had a typical resistivity of <20 Ω per μm of their length, roughly independent of *d*. The contacts exhibited linear I-V characteristics without any evidence of a Schottky barrier down to nV biases and liquid-helium temperatures. Our devices could sustain very large current densities of up to $\approx 10^8$ A/cm$^2$. We emphasize that despite many steps involved in the film and device microfabrication, the described procedures were found to be highly reproducible, except for the peeling-off process that in the end relies on manual selection of a most suitable thin flake.

FIGURE CAPTIONS

Figure 1. Graphene films. Optical micrograph of a rather large uniform graphene film obtained by repeated peeling of HOPG. The inset shows an atomic-force-microscope (AFM) image of a µm- sized area of this 3-nm film near its edge.

Figure 2. Field effect in graphene films. Main panel: a typical dependence of graphene film's resistance on gate voltage at room temperature; $d \approx 5$nm. $R_0 \approx 100\Omega/\square$ is the resistivity at zero $V_g$. **a**, Schematic view of our experimental devices. **b**, Optical micrograph of one of them. The horizontal wire has a width of 5 µm. **c**, The expected behaviour for a 5-nm film of graphite. For simplicity, the calculations assumed $\mu_e = \mu_h$. The dashed line at zero $V_g$ indicates the compensated state $n_h = n_e$ while the blue and red shaded areas show the regions of entirely electron and hole conductances, respectively.

Figure 3. Influence of electric-field doping on the Hall effect. The main panel shows the dependence of Hall coefficient $R_H$ on gate voltage $V_g$ for the graphene film of Fig. 2. Solid curves indicate the regions of hole, mixed and electron conductivities (from left to right). **a**, The calculated dependence $R_H(V_g)$; notations as in Fig. 2c. **b**, Even for this relatively thick graphite film ($d \approx 50$nm), Hall resistance $R_{xy}(B)$ strongly depends on gate voltage, although the changes are limited to a small region of $n < n_i$, so that both electrons and holes remain present. From top to bottom, the plotted curves correspond to $V_g$ = -90, -30, 0, 30 and 90 V. At zero $V_g$, $R_{xy}(B)$ curve is flat indicating the case of a compensated semimetal while negative (positive) gate voltages induce a large positive (negative) Hall effect. **c**, Changes in Hall coefficient for the same device as in (b).

Figure 4. Shubnikov-de Haas oscillations in graphene films. The main panel shows ShdH oscillations in a 5-nm film for three different gate voltages; temperature 3K. **a**, ShdH oscillations with varying $V_g$ in a fixed magnetic field of 6T at 0.3 K. For clarity, a smooth monotonic background has been subtracted. **b**, As a reference, we show the behaviour of $R_{xx}(B)$ and $R_{xy}(B)$ in a thick film of HOPG ($d \approx 8\mu m$) at 0.3 K. Plateau-like features in $R_{xy}$ with characteristic overshoots are common for such thick samples but disappear in nm-thin films, probably because of their lower μ. The plateaux occur around resistance values of $(h/2e^2)/i\nu$ where $h/2e^2$ is the resistance quantum, $\nu$ the number of occupied Landau levels ($\nu = 1$ at $B \approx 4T$) and $i$ the number of graphene layers in the film. This observation offers further support for recent speculations about a possible quantum Hall effect behaviour in graphite [9,15].

# LIST OF REFERENCES

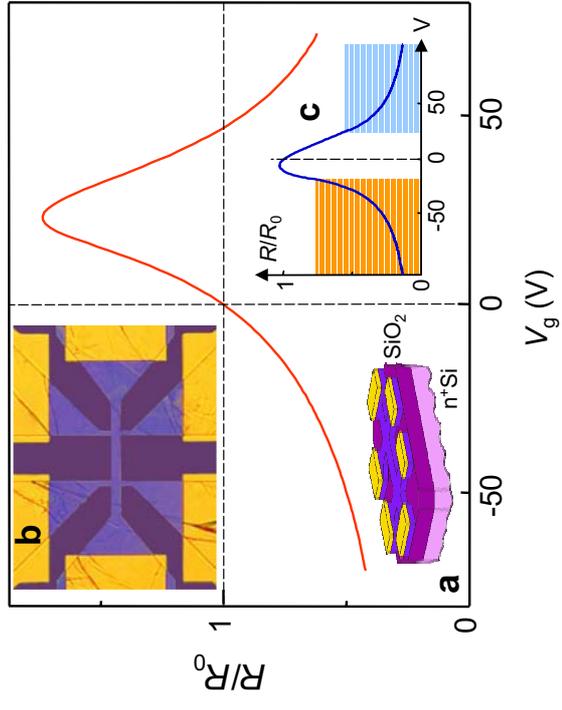

Figure 2

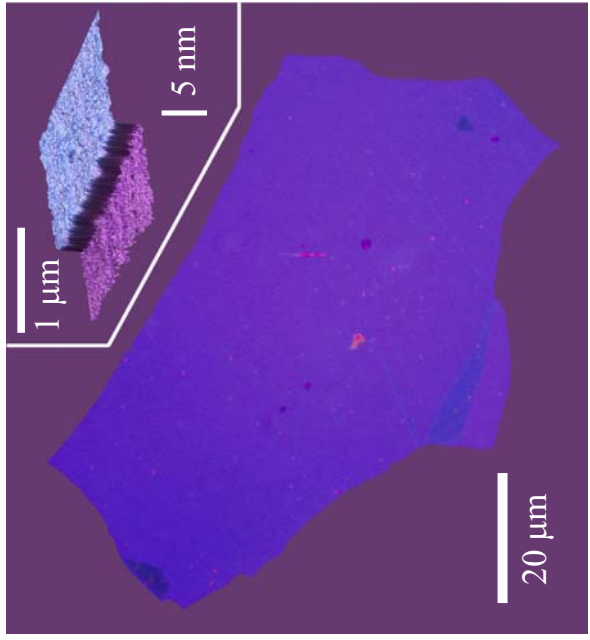

Figure 1

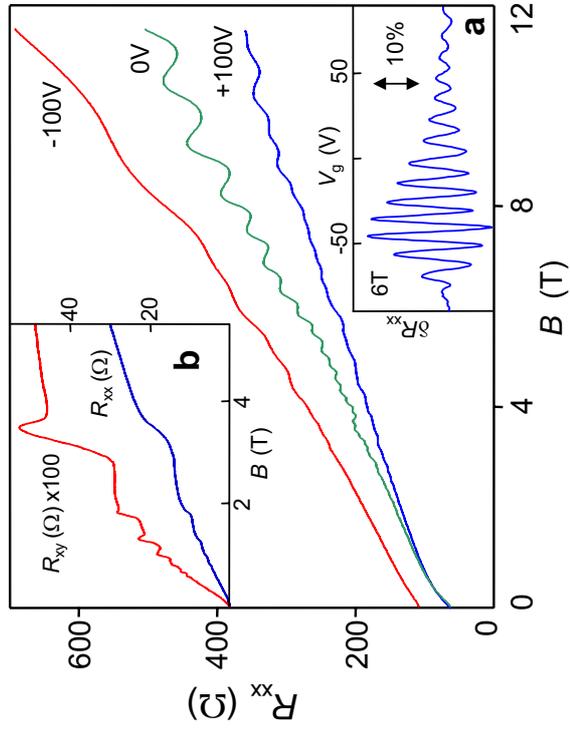

Figure 4

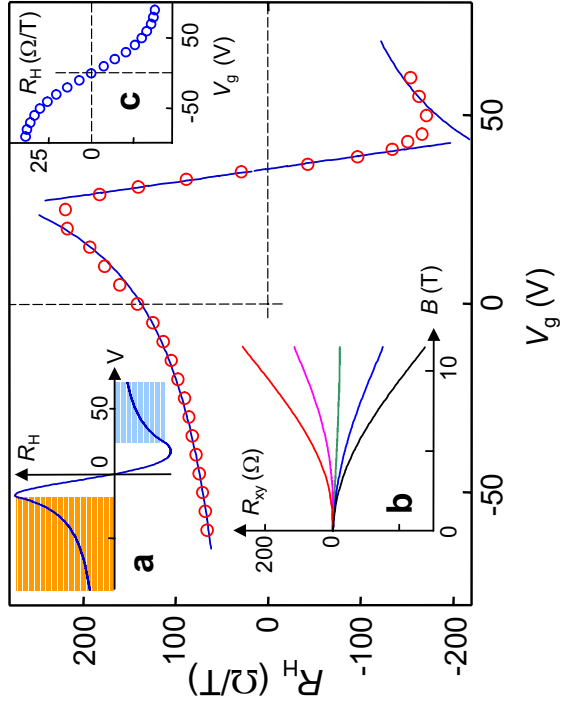

Figure 3